\documentclass[12pt,a4paper]{article}
\usepackage{amssymb}
\usepackage{amsmath}
\usepackage{pst-node}
\newcommand{\ud}{\mathrm{d}}

\begin{document}
\title{\textbf{Deterministic definition of the capital risk}}
\author{Anna Szczypi\'nska\\ Institute of Physics, University of Silesia, \\ Uniwersytecka
4, Pl 40007 Katowice, Poland \\ e-mail:
zanna12@wp.pl\\
Edward W. Piotrowski\\ Institute of Mathematics,
University of Bia\l ystok,\\ Lipowa 41, Pl 15424 Bia\l ystok,
Poland\\ e-mail: ep@alpha.uwb.edu.pl
}
\date{}
\maketitle
\def\R{{\mathbf R}}
\def\p{{\mathbf p}}
\def\f{{\mathbf f}}
\def\U{{\mathbf U}}

\begin{abstract}
In this paper we propose a look at the capital risk problem inspired by deterministic, known from classical mechanics, problem of juggling. We propose capital equivalents to the Newton's laws of motion and on this basis we determine the most secure form of credit repayment with regard to maximisation of profit. Then we extend the Newton's laws to models in linear spaces of arbitrary dimension with the help of matrix rates of return. The matrix rates describe the evolution of multidimensional capital and they are sensitive to both quantitative changes of individual elements and flows between them. This allows us for simultaneous analysis of evolution of complex capital in both continuous and discrete time models.
\end{abstract}

\section{Introduction}
At first sight, physics and economics are completely different in nature. In economics, in contrast to physics, we do not know any inviolable mathematical laws of market---the basic universal laws as for which everybody is unanimous. Therefore, even the most complex economic models are incomplete, that is, they are not capable of giving any suitable predictions about the future. John Kay from London Business School said that ''Economic forecasters$\ldots$ they all say more or less the same thing at the same time. And what they say is almost always wrong. The differences between forecasts are trivial relative to the differences between all forecasts and what
happens.'' \cite{kay}. That is why, physicists are attempting to compete with economists in understanding and explaining economic phenomena. This undertaking is called econophysics. Of course, physics make use of tools which cannot be directly applied to economic phenomena. On the other hand, it could be surprising if some phenomena, well understood in physics, did not appear in another form in economics \cite{ball}. In this paper, we show similarity between classical mechanics of material points and modelling of credits. Modern definition of the financial risk is generally connected with probabilistic methods of prices motion modelling in capital markets. We propose a different look at the capital risk problem inspired by deterministic problem\footnote{D. Dobija and M. Dobija in paper \cite{dobija} pay attention to existence the deterministic risk premium in the economic exchange.} of juggling, see Ref.~\cite{ryzkap}. On this basis we determine the most secure form of credit repayment with regard to maximisation of profit. Then, we extend this rule to models in linear spaces of arbitrary dimension with the help of matrix rates of return \cite{matr}. The matrix rates describe portfolios of arbitrary type and extend portfolio analysis to the complex variable domain. It is worth to signal the broad perspectives of application of such deterministic variant of the financial risk.

  %We show an isomorphism between classical mechanics of material points and modelling of credits, and on this basis we determine the most secure form of credit repayment with regard to maximisation of profit. Then, we extend this rule to models in linear spaces of arbitrary dimension with the help of matrix rates of return. The matrix rates describe portfolios of arbitrary type and extend portfolio analysis to the complex variable domain. This allows us for simultaneous analysis of evolution of complex capital in both continuous and discrete time models. It is worth to signal broad perspectives of application of such deterministic variant of the financial risk.

\section{Problem of juggling}
Let us consider the following problem:\textit{ A man needs to carry balls across a bridge. The bridge will collapse under the weight greater than that of the man plus a ball. How should the man get the balls across the bridge without breaking it, if he has only one chance?} Juggling, is it a good idea? From the Newton's laws of motion follows that (considering a vertical component), the rate of change of momentum of the carried balls, with the total weight $Q$, is equal $$\ud p(t)=(Q-f(t))\ud t \,,$$
where $f(t)$ is the man's reaction force on the balls, and the bridge on the balls as well.
Both before and after being carried, the balls are laying on the ground, that is, their momentum does not change as a result of this operation:\vspace{-0.5em}
$$p(T)-p(0)=\int_{p(0)}^{p(T)}\negthinspace\ud p(t)=\int_{0}^{T}\negthinspace\negthinspace(Q-f(t))\ud t=0\, ,$$
where $T$ is the time of the bridge crossing. From the above equation we obtain 
$$\frac{1}{T}\int_{0}^{T}\negthinspace\negthinspace f(t)\ud t=Q\,,$$
thus, the average pressure of the carried balls on the bridge in period $T$ is equal theirs weight!\\
In order to transfer the balls across the bridge, the largest value of the pressure force on the bridge must be smaller than the minimum force which causes collapse of the bridge. For different ways of transport which are uniquely defined by the functions $f(t)$, the collapse risk of the bridge can be measured with the help of $f_{max}$. A natural metric, measuring a distance (level of risk) between two transport ways of the balls, is an absolute value of the difference of values $f_{max}$. From the Mean Value Theorem we obtain $$\frac{1}{T}\int_{0}^{T}\negthinspace\negthinspace f(t)\ud t\leqslant f_{max} :=\max_{0\leq t\leq T}\left[f(t)\right]\,,$$
so the minimum maximum ($\rm{minmax}$) value of this pressure while the bridge crossing is equal to the total weight of the balls. Therefore, a resignation from juggling is the best method of effective transport of the balls---then $f_{max}$ is minimal. Juggling increases only the maximal value of the man's pressure force on the bridge, what, in the best case, enlarges the collapse risk of the bridge.
%Let us notice that solving problem of juggling, we took under attention all possible ways of juggling because the solution does not depend on the function $f(t)$.

\section{Capital equivalents to the Newton's laws of motion}

Let usury be any process of capital transfer, in return of that, a borrower undertakes to make at later dates payments of the capital to a lender, which in his opinion should justify the loan. This process is a compulsion-free market service. Both the repayment of the capital and the loan could have a statistical character. Then, bill of exchange, credit, bond, stock, futures contract, option, insurance policy will be the usury in this sense. We show that there is an isomorphism between the usury and the problem of juggling. \\
Let $p_\tau (t)$ be the usurer's account balance (or debt), that is, the capital assets at the moment $t$, expressed in any capital units at the fixed past (or future) moment $\tau$; and $f_{\tau}(t)$ be the intensity of capital flow on (or of) an account discounted to the moment $\tau$. Then, capital equivalent to the Newton's second law of motion we can formulate as follows: \textit{The velocity of changes of an account balance\footnote{In economics the flux of capital.} is equal to the intensity of capital flows through it:}
$$\frac{\ud p_\tau (t)}{\ud t}=f_\tau (t)\,.$$
A consequence of this formula is insensitivity of description of the capital flows to the correction of the account balances by a constant. Usually the above equation is used in the integral form:
$$p_\tau (t_2)-p_\tau (t_1)=\int_{p_\tau (t_1)}^{p_\tau (t_2)}\negthinspace\negthinspace \ud p_\tau (t) =\int_{t_1}^{t_2}\negthinspace\negthinspace f_\tau (t)\ud t\,,$$ because we calculate in capital units but not in theirs intensities. Let us notice that, this is the law of conservation of the capital, because the difference in an account balance is always the result of balance of the capital flows, which occur in this account.\\
Capital equivalent to the Newton's first law of motion is following: \textit{An account balance does not change if the intensity of the capital flows equals zero $f_\tau (t)\equiv 0$.} Like in physics, this law we interpret as the postulate about existence of proper functions of utility which are equivalents to inertial reference frames.\\
The Newton's third law of motion in capital language is: \textit{The intensity of the capital $f_{B\mapsto A}$ influencing the person's $A$ account, which comes from the person $B$, must be balanced by the intensity $f_{A\mapsto B}$ coming from the person $A$ and influencing the person's $B$ account, that is}\vspace{-0.5em}
$$f_{A\mapsto B}+f_{B\mapsto A}=0\,.$$ The above law in financial markets implies the non-existence of free lunch phenomenon. In economics, it is enough if expected values of intensities of the capital flows fulfil the third law \cite{ryzkap}. This mean that somebody can eat free lunch, but someone else must ever pay for this lunch.\\

We showed an isomorphism between the usury and the problem of juggling. On this basis we now calculate the most secure form of the usury repayment. First, to compare values of a good at different moments, we define the utility function of the capital. The usurer know, how much (at most $U_0(\tau, t)$) he is able to lend in the units $\$_{\tau}$ at the moment $\tau$, so that, at later (or earlier) moment $t$ get back a unit $\$_{t}$:
$$ utility(\tau,\$_{t})=U_0(\tau, t)\,\$_{\tau}\,.$$ 
The function $U_0(\tau, t)$ is the utility of a unit of money for the usurer. To see properties of the function $U(\tau, t)$, see Ref.~\cite{algkr}. Let us assume that the loaned and returned amounts are equally useful for the usurer. We change the reference frame---this is $alias$ convention.\\
Let us consider a discrete model of the usury. Then, the intensity of the capital flow is equal\vspace{-0.5em}
$$f_\tau (t)=\sum_{k=0}^{N}\varphi_{\tau}(k)\,\delta (t-t_k)\,,$$
where $\varphi_{\tau}(k)=U(\tau, t_{k})\varphi(k)$ is the amount of instalment discounted to the moment $\tau$, $\varphi(k)$ is the face value of the $k$-th instalment of repayment, $\delta$ is the Dirac delta, and $N$ is number of instalments. Let $\varphi_\tau(0)$ be an amount of the loan and a negative number, and $\varphi_\tau(k)$, for $k=1,\ldots ,N$, be the successive instalments of repayment discounted to the moment $\tau$. The loan is balanced from the usurer point of view, so that 
$$p_\tau (T)-p_\tau (0)=\int_{0}^{T}\negthinspace\negthinspace f_\tau (t)\ud t=\sum_{k=0}^{N}\varphi_{\tau}(k)=0\,.$$
The analogy of the weight $Q(t)$, there are any intensities of the capital flows through the usurer's account, which are not included in the contract of the usury. Without loss of generality, we assume that $Q(t)=0$. Let us assume that the usurer's preference are $\rm{minmax}$---he strives for such contract, to the maximum possible loss if the borrower default on the credit payment, will be minimum. We knowingly assumed such model of minimising risk because minimising average risk in some situations is insufficient, for example, when the borrower takes such strategy of the credit repayment that one of the instalments be a very large in relation to the rest of instalments. From the Mean Value Theorem the minimum loss occurs for all equally profitable periods:\vspace{-0.5em}
$$\varphi_\tau (k)=const_\tau =\frac{1}{N}\sum_{m=1}^{N}\varphi_{\tau}(m)=\frac{1}{N}\sum_{m=1}^{N}U(\tau, t_m)\,\varphi(m)\,,$$ 
for $k=1,\ldots ,N$. This is the usury with the minimal risk for the usurer. This is a credit with really fixed instalments of repayment. Dual to this, there is a credit with nominally fixed capital instalments---that is, the most popular credit from among all kinds of credits.\\
How about juggling---adopting different present values for the successive instalments of repayment? Juggling in the case of a loan repayment would be adopting different present values for the successive instalments of repayment. As a natural risk measure of any usury contract, when the usurer's preferences are $\rm{minmax}$, we can use the average of the squared differences between the sequence of repayments $\left\lbrace \varphi_\tau (k)\right\rbrace $ and its mean \cite{stein}\footnote{In this paper Steinhaus explain in which way a change of preferences influences definition of proper risk measure, so it is possible to define the risk measure for another preferences of the usurer.}. Such measure guarantees that the safest way is the loan repayment without juggling, because it has a unique minimum at the mean!

\section{Utilities of the capital and rates of return}
%\subsection{A continuous time model}
The usurer's account balances $p_\tau (t)$ and $p_\tau (t')$ at the moment $t$ and $t'$ respectively, expressed in any capital units at the moment $\tau$, fulfil the equation $p_\tau (t)=U(t, t')\,p_\tau (t')\,.$
We define $r(t,t'):=\ln U(t,t')$ as the range rate of return. The utility and the range rate of return are given by the formulas:
$$U(t,t')={\rm e}^{\int_{t'}^{t}r(t'')\ud t''}\,,\;\;\;\;\;\;\;r(t,t')= {\int_{t'}^{t}r(t'')\ud t''}\,,$$
where $r(t)$ is the temporary (differential) rate of return. If we know $r(t)$ and $p_\tau(t')$, we can interpret\vspace{-0.5em}
$$p_\tau (t)={\rm e}^{\int_{t'}^{t}r(t'')\ud t''} p_\tau(t')$$ as the temporary account balance. From this equation, we obtain that the velocity of changes of an account balance is equal
$$\frac{\ud p_\tau (t)}{\ud t}=r(t)\,p_\tau (t)\,.$$ From the Newton's second law of motion, we know that the velocity of changes of an account balance is equal to the intensity of the capital flows through this account: $f_\tau (t)=r(t)\,p_\tau (t)\,.$
In the integral form, we obtain that the usurer's profit is equal
$$p_\tau (t)-p_\tau(t')=\int_{t'}^{t}r(t'')\,p_\tau(t'')\ud t''\,.$$ 

%If $r(t)=r=const$, then $r(t,t')=(t-t') r\,,$ $U(t,t')={\rm e}^{(t-t') r}\,$, $f_\tau (t)=r~p_\tau (t)\,,$ and the profit equals\vspace{-0.5em} $$p_\tau (t)-p_\tau(t')=r \int_{t'}^{t}p_\tau(t'')\ud t''\,.$$

\subsection{A discrete time model}
If we want to calculate a value of the capital in a discrete time model, we have to replace a differential equation with a difference equation. The lower rate of interest is defined as: $\underline{r}(t,t'):=U(t,t')-1\,$ and it fulfils the following difference equation:$$p_\tau(t)-p_\tau(t')=\underline{r}(t,t')\,p_\tau(t')\,.$$
The upper rate of interest is defined as: $\overline{r}(t,t'):=1-U(t,t')^{-1}\,$ and it fulfils the equation: $$p_\tau(t)-p_\tau(t')=\overline{r}(t,t')\,p_\tau(t)\,.$$ 
Comparing above formulas, we obtain the relation between both types of rates of interests:
$$(1+\underline{r}(t,t'))(1-\overline{r}(t,t'))=1\,.$$ Then, the account balance $p_\tau(t)$ at the moment $t$, expressed in capital units at the moment $\tau$, fulfils the equations:
$$p_\tau(t)\negthinspace=\negthinspace p_\tau(t')\negthinspace\prod_{k=t'}^{t-1}\negthinspace\left( 1\negthinspace+\negthinspace\underline{r}(k,\negthinspace k\negthinspace+\negthinspace1) \right) \,,\;\;\;\;p_\tau(t)\negthinspace=\negthinspace p_\tau(t')\negthinspace \prod_{k=t'}^{t-1}\negthinspace\left(1\negthinspace-\negthinspace\overline{r}(k,k\negthinspace+\negthinspace1)\right)^{-1}\,.$$
With the help of these formulas, we can calculate the value of the capital at any discrete time moments, if we know the lower and upper rates of interest. 

\section{Generalisation of the usury on linear spaces of arbitrary dimension}

Of course it does not matter, if the account balance and intensity relate to one or more goods. The Newton's laws are fulfilled for models in linear spaces of arbitrary dimension. Let us consider intensity of the capital flows by $n$ usurer's accounts at the moment $t$ discounted to $\tau$. It is given by the formula:
$$\frac{\ud \p_\tau (t)}{\ud t}=\f_\tau (t)=\R(t)\,\p_\tau (t)\,,$$
where $\p_\tau (t)\in \mathbb{R}^n$ is the balance of $n$ accounts at the moment $t$ expressed in capital units at the moment $\tau$ and $\R(t)$--a real matrix of dimension $n\times n$, is the temporary (differential) matrix rate of return \cite{matr}. The formal solution of the above equation is given by the formula:
$$\p_\tau (t) = \left( \mathcal{T} {\rm e}^{\int_{t'}^{t}\negthinspace\R (t''\negthinspace)\ud t''}\right)\p_\tau(t')\,,$$
where $\mathcal{T}$ denotes the chronological ordering operator. The chronologically ordered exponential function is equivalent to the utility of the capital:
$$\U(t, t'):=\mathcal{T}
\rm{e}^{\int_{t'}^{t}\negthinspace\mathbf{R} (t^{''}\negthinspace)
\ud t^{''}}\negthinspace\negthinspace=\mathbf{I}+\negthinspace\int_{t'}^{t}\negthinspace\mathbf{R} (t'_1 ) \ud t'_1
+\int_{t'}^{t}\negthinspace\mathbf{R} (t'_1 ) \negthinspace\int_{t'}^{t'_1}\negthinspace\negthinspace\mathbf{R}(t'_2 )\ud t'_2 \ud t'_1 +\ldots \,,$$
where $\mathbf{I}$ is the unit matrix. The above expression is the special case of the Volterra series, see Ref.~\cite{volter}. The matrix rates of return take into account not only quantitative changes of individual accounts, but also flows between them. In \cite{matr} it is shown that the flows of capital can be recorded even if no decisions about capital operations are taken. Such situations require the matrix description. 

%If the differential matrix rate is constant $\R(t)=\R$, then $\mathcal{T}$ is the identity and $\U(t,t')=\rm{e}^{(t-t')\mathbf{R}}.$

\subsection{A discrete time model}
Similarly as in the case of one dimension, we can define the matrix lower rate $\underline{\R}(t,t')$ and the matrix upper rate $\overline{\R}(t,t')$ which fulfil the linear homogeneous difference equations:
\begin{equation}
\p_\tau(t)-\p_\tau(t')=\underline{\R}(t,t')\,\p_\tau(t')\,,\;\;\;\;\;\;\;\;\;\p_\tau(t)-\p_\tau(t')=\overline{\R}(t,t')\,\p_\tau(t)\,.\label{rown}
\end{equation}
Comparing the above formulas, we obtain the relation between both types of matrix rates\footnote{If the matrices $\mathbf{I}+\underline{\R}(t,t')$ and $\mathbf{I}-\overline{\R}(t,t')$ are nonsingular.}:
$$(\mathbf{I}+\underline{\R}(t,t'))(\mathbf{I}-\overline{\R}(t,t'))=\mathbf{I}\,.$$  Then, the solutions of the equations (\ref{rown}) take the forms respectively:
$$\p_\tau(t)=\left( \mathcal{T}\negthinspace\prod_{k=t'}^{t-1}\negthinspace\left( \mathbf{I}+\underline{\R}(k,\negthinspace k\negthinspace+\negthinspace 1) \right)\right)\p_\tau (t')\,,$$
 $$\p_\tau (t')=\left(  \mathcal{T'}\negthinspace \prod_{k=t'}^{t-1}\negthinspace\left(\mathbf{I}-\overline{\R}(k,k\negthinspace+\negthinspace 1)\right)\right)\p_\tau(t) \,,$$ where $\mathcal{T}^{'}$ is the antichronological operator. If the matrix rates at different moments of time commute, we can neglect the operators $\mathcal{T}$ and $\mathcal{T'}$. 

\section{Conclusions}
Exploration by looking for analogies has always been a fundamental method of research. At the time when communication methods are advanced, this method is particularly encouraging. There may be multiple benefits in grasping of similarity of models which explain a nature of distant phenomena. The problem of juggling and the analogous problem of the most secure usury show that risk does not necessarily involve random phenomena---it can have a pure deterministic nature. Such variant of the capital risk can be applied to the capital insurances, see Ref.~\cite{ryzkap}. The proposed multidimensional extension of the Newton's law can be particularly useful. This description does not depend on the choice of basic goods. Therefore, we can consider the complex extended space of goods and the basis of complex eigenvectors of the matrix rates in which the evolution of every multidimensional capital can be represented as a set of noninteracting complex capital investments. In this complex basis we do not observe any flows of the capital, but the autonomous growth of individual components only. Then a description of the capital evolution is the most easily \cite{matr}.

\end{document}